\documentclass{icrc29}
\usepackage{graphicx,amssymb,amsmath,times}
\begin{document}
%Title of paper
\title[]{Using transport coefficients of cosmic rays in turbulent
magnetic fields to determine hybrid viscosity in hot accretion disks
around AGN}

\author[P. Subramanian et al.] {Prasad Subramanian$^{a}$
        \newauthor
Peter A. Becker$^b$, Menas Kafatos$^b$ \\
        (a) IUCAA, P. O. Bag 4, Ganeshkhind, Pune - 411007, India\\
        (b) School of Computational Sciences, George Mason University, Fairfax, VA 22030, USA
        }
\presenter{Presenter: Prasad Subramanian (psubrama@iucaa.ernet.in), \  
ind-subramanian-P-abs1-og23-oral}

\maketitle

\begin{abstract}

The nature of the viscosity operative in hot, two-temperature accretion
disks around AGN has been a long-standing, unsolved problem. It has been
previously suggested that protons, in conjunction with the turbulent
magnetic field that is likely to exist in the accretion disk, might be
crucial in providing this viscosity. Several authors have recently
determined diffusion coefficients for charged particles (cosmic rays)
propagating in turbulent magnetic fields by means of extensive Monte Carlo
simulations. We use the diffusion coefficients for protons determined by
these simulations to find the effective mean free path for protons in hot
accretion disks. This in turn yields good estimates of the viscosity due
to energetic protons embedded in the turbulent magnetic field of a hot,
two-temperature accretion disk. We combine this with a simple
two-temperature accretion disk model to determine the Shakura-Sunyaev
$\alpha$ viscosity parameter arising out of this mechanism. We find that
 protons diffusing in the turbulent magentic field embedded in a hot accretion disk provide a
physically reasonable source of viscosity in hot accretion disks around
AGN.

\end{abstract}

\section{Introduction}

The microphysical viscosity mechanism operative in hot, two-temperature 
accretion disks around AGN has been 
a subject of intense research, and is a yet-unsolved question. 
In the steady state, such a viscosity mechanism enables matter around
the central black hole to lose its angular momentum and accrete onto it. Recent advances in the theory
of the magnetorotational instability in cold accretion disks (Balbus \& Hawley 1998 and references therein) 
suggest the existence of a chaotic, tangled magnetic field,
together with a relatively well-ordered large-scale toroidal field embedded in the accretion disk around
a black hole. The ``magnetic viscosity'' arising from stresses related with the tangled magnetic fields 
is thought to be the operative viscosity mechanism in cold accretion disks. However, in hot, two-temperature
accretion disks, where the protons are hot ($T_{i} \sim 10^{10}$--$10^{12}$K) 
and their mean-free path is comparable to (or larger than)
the scale height of the disk, this picture is inadequate. The theory of MRI-like instabilities under conditions
similar to those prevalent in hot accretion disks is still 
in its infancy (Balbus 2004; Quataert et al. 2002; Sharma et al. 2003).
However, these initial developments do suggest that such an instability will also culminate in a tangled
magnetic field embedded in the accretion disk. 
Assuming the presence of such a tangled magnetic field in a hot accretion disk,
Subramanian, Becker \& Kafatos (1996) (SBK96 from now on) 
showed that the operative viscosity mechanism was a ``hybrid'' one,
arising from hot protons colliding with kinks in magnetic field lines. 
SBK96's treatment had 
$\xi \equiv \lambda_{\rm coh}/H$, as a free parameter, where $\lambda_{\rm coh}$ is
the coherence length of the magnetic field and $H$ is the disk scale height. The quantity $\xi$ thus provides
a measure of the extent to which the magnetic field is tangled, in comparison with the disk height. 
SBK96 were only able to
treat one value of $\xi$ at a time, and could not consider the overall effect arising from
the presence of a spectrum of $\xi$s (i.e., a spectrum of scales over which the field is tangled).
Furthermore, they did not consider the presence of an ordered, large-scale toroidal field, which is
suggested by MRI simulations, albeit in cold accretion disks. In the present treatment, we use results from
extensive Monte-Carlo simulations of cosmic rays propagating through tangled magnetic fields published 
in the literature in order to remedy these shortcomings and arrive at a more rigorous estimate of hybrid viscosity.

\section{Cosmic ray transport through turbulent magnetic fields}
Several authors have studied the diffusion and drift of charged particles across turbulent magnetic fields.
While analytical perturbative studies in the low-turbulence regime have been carried out for a long 
time (e.g., Parker 1965; Jokipii 1966)
numerical experiments suitable to the high turbulence regime are relatively recent (Giacalone \&
Jokipii 1999; Casse et al 2002;
Candia \& Roulet 2004). There is
a large scale magnetic field typically assumed to be present, in addition to the turbulent conponent, and 
such studies strive to characterize
the motion of cosmic rays through turbulent magnetic fields by means of diffusion coefficients perpendicular and
parallel to this large scale field. We follow the treatment of Candia \& Roulet (2004), who have performed
Monte Carlo simulations following cosmic ray trajectories in the presence of a tangled magnetic field for
a variety of turbulence levels and magnetic rigidities. They provide convenient analytical fits to the parallel and
perpendicular diffusion coefficients arising from their simulations. 
\section{Hot protons diffusing through tangled magnetic fields: hybrid viscosity}
We use the analytical fits to the diffusion coefficients 
provided by Candia \& Roulet (2004) to compute 
the viscosity arising out of hot protons diffusing through tangled magnetic fields.
The advantage of this approach is that it takes the interaction of the hot protons with all the spatial
scales of turbulent magnetic fields, and not only with one scale, as in SBK96. 
As suggested by simulations of the magnetorotational instability in
cold accretion disks, we presume that there exists a large-scale magnetic
field in the toroidal direction embedded in the accretion disk, in addition to a tangled field. We are 
interested in diffusion coefficients (and consequent viscous angular momentum transport) in the radial
direction, which is perpendicular to this large-scale toroidal magnetic field. The perpendicular
diffusion coefficient is, by definition, the one which characterizes motion perpendicular to the large scale,
ordered magnetic field.We therefore use 
expressions for the perpendicular diffusion coefficient $D_{\perp}$ provided by Candia \& Roulet (2004).

From Eqs (18) and (19) of Candia \& Roulet (2004), adopting the low rigidity assumption ($\rho \leq 0.2$) 
for simplicity, we get
{\begin{eqnarray}
\nonumber
D_{\perp} = v_{\rm rms}\,H\,D_{c}\,\,\,\,\,\,{\rm cm^{2}\,s^{-1}}\, , \\
D_{c} = N_{\perp}\,(\sigma^2)^{a_\perp}\,\rho\,\frac{N_{||}}{\sigma^{2}} \,
\biggl ( \biggl (\frac{\rho}{\rho_{||}}\biggr )^{2(1-\gamma)} + \biggl (\frac{\rho}{\rho_{||}}\biggr )^{2} \biggr )^{1/2} \, ,
\label{eq1}
\end{eqnarray}}
where we have used the disk height $H$ for the maximum scale length $L_{\rm max}$. The quantity
$v_{\rm rms} \equiv \sqrt{3\,k_{B}\,T_{i}/m_{i}}$ (where $k_{B}$ is Boltzmann's constant, 
$T_{i}$ is the proton temperature and $m_{i}$ denotes the proton mass) is the proton rms velocity.
We use $v_{\rm rms}$ instead of the speed of light in defining $D_{\perp}$ as Candia \& Roulet (2004) do,
since the hot ($T_{i} \sim 10^{10}$K) protons we consider have energies of the order of 1 Mev, and 
are therefore non-relativistic. Candia \& Roulet's (2004) results are quite valid for protons in this
energy range too, unless the proton thermal velocities are close to the Alfven speed, in which case
coupling to Alfven modes needs to be taken into account. Although simulations of the nonlinear stage
of the MRI in hot accretion disks have not yet been performed, corresponding simulations for
conditions corresponding to cold disks reveal that the final plasma $\beta$ saturates at values $\sim 500$.
Under these conditions, the particle thermal velocities are far in excess of the Alfven velocity.
The quantity $\rho$ denotes the rigidity, which is defined as the ratio of the Larmor radius to
the maximum scale length. The quantity $\sigma^{2}$ denotes the ratio of the energy density in the turbulent
magnetic field to that in the large scale magnetic field. The
values of the parameters $a_\perp$, $N_{||}$, $,N_{\perp}$ are given in Table 1 of
Candia \& Roulet (2004) for various kinds of turbulence. For the sake of concreteness, we use
values corresponding to the first line of that table; i.e., Kraichnan turbulence.

The coefficient of dynamic viscosity 
is usually defined as $\eta \,({\rm g\,cm^{-1}\,s^{-1}})\, = N m v \lambda$, 
where $N$ and $m$ are the density and
mass of the relevant particles, $v$ is the relevant velocity (usually the thermal rms velocity) and
$\lambda$ is the relevant mean free path (Spitzer 1962; Mihalas \& Mihalas 1984). The diffusion co-efficient
$D$ is usually defined as $D \,({\rm cm^{2}\,s^{-1}})\, = v \lambda$, where $v$ is the relevant velocity and
$\lambda$ is the relevant mean free path. Combining these two expressions, we get $\eta = N m D$.
The coefficient of dynamic viscosity arising out of the radial transport of hot protons (i.e., perpendicular to the large scale toridal field) can therefore be written as
\begin{equation}
\eta_{\rm hyb} = N_{i}\,m_{p}\,D_{\perp} = 7.54\times 10^{10}\,D_{c}\,\tau_{es} \,\,\,\,\,{\rm g\,cm^{-1}\,s^{-1}}\, ,
\label{eq3}
\end{equation}
where we have used Eq~(\ref{eq1}) and Eq. (A6) of SBK96. The quantity $N_{i}$ is the proton number density and
$m_{i}$ is the mass of a proton.

Using Eq~(\ref{eq3}) and Eqs. (A1), (A4), (A6), (A9) of SBK96 in Eq. (3.4) of the same paper, we get
\begin{equation}
1 + \tau_{es} = 3.07 \times 10^{-3} \,\biggl (\frac{\dot{M_{*}}}{M_{8}} \biggr )^{-1} \,f_{1}^{-1/2}\,f_{2}^{-1}
\,y\,R_{*}^{3/2}\,D_{c} \, 
\label{eq4}
\end{equation}
and
\begin{equation}
\alpha_{\rm hyb} = 2.557\,f_{1}^{-1/2}\,D_{c} \, .
\label{eq5}
\end{equation}
We have used the same notations as in SBK96; $\tau_{es}$ denotes the electron scattering optical depth,
$\alpha_{\rm hyb}$ is the Shakura-Sunyaev viscosity parameter (Shakura \& Sunyaev 1973) due to
protons diffusing through tangled magnetic fields, $\dot{M}_{*}$ is
the accretion rate in units of solar masses per year, $M_{8}$ is the black hole mass in units of $10^{8}$ solar
masses and $f_{1}$, $f_{2}$ and $f_{3}$ are the relativistic correction factors used in SBK96 
and references therein.
The results for $\alpha_{\rm hyb}$ and $\tau_{es}$ from Eqs~(\ref{eq5}) and (\ref{eq4}) can be used in 
Eqs. (A7)--(A9) of SBK96 to determine the disk structure as a function of radius for a given accretion rate 
$\frac{\dot{M_{*}}}{M_{8}}$. For a given set of parameters characterizing the
turbulence (which yields a given value of $D_c$ via Eq.~\ref{eq1}), Eq~(\ref{eq4}) shows that the 
requirement $\tau_{es} > 0$
implies an upper bound on the accretion rate $\frac{\dot{M_{*}}}{M_{8}}$.

In Figure 1, we have shown the results for the following set of parameters characterizing the
turbulence: $\gamma = 3/2$, $N_{||} = 2$, 
$\rho_{||} = 0.22$, $N_{\perp} = 0.019$, $a_{\perp} = 1.37$. These values 
are taken from the first line of Table 1 of Candia \& Roulet 2004, and characeterize a Kraichnan turbulent spectrum.
The value of the magnetic rigidity $\rho$ is taken to be 0.2 and the parameter characterizing the level of
turbulence, $\sigma^{2}$, is taken to be 0.1. In other words, the proton larmor radius is 0.2 times the disk height,
and the energy in the turbulent magnetic fields is 0.1 times that in the ordered magnetic field. The accretion
rate $\frac{\dot{M_{*}}}{M_{8}} = 1.6 \times 10^{-4}$, which is close to the allowed upper limit above which
the optical depth $tau_{es}$ will become negative. This value of $\frac{\dot{M_{*}}}{M_{8}}$ corresponds to
$\frac{\dot{M}}{\dot{M}_{\rm E}} = 7.27 \times 10^{-4}$, where $\dot{M}_{\rm E}$ is the Eddington accretion rate.
The solid line shows the value of $\alpha_{\rm hyb}$, while the dotted line shows the quantity $0.1\,H/R$. The
dashed line shows the proton temperature in units of $10^{12}$K. We have found that the values of $\alpha_{\rm hyb}$
are in the range of $0.01--0.02$ for $\sigma^{2} = 0.5$.

\begin{figure}
\begin{center}
\includegraphics[width=3.5in]{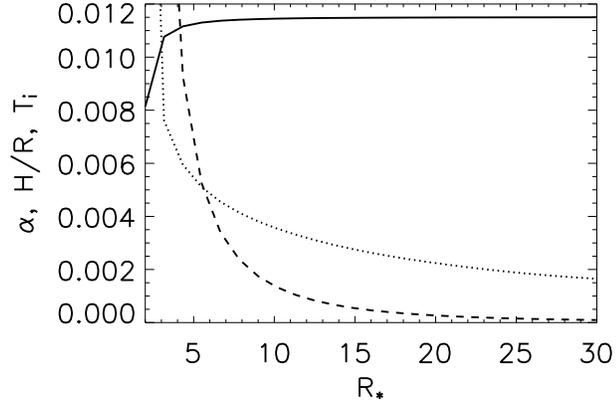}
\end{center}
\caption{Results corresponding to an accretion rate of $\dot{M}/\dot{M_{\rm E}} = 7.27 \times 10^{-4}$. 
The solid line is
a plot of the $\alpha$ parameter, while the dotted line shows the quantity $0.1\,H/R$. The dashed line shows the
the proton temperature $T_{i}$ in units of $10^{12}$K.}
\end{figure}
\section{Conclusions}
We have used results from extensive Monte Carlo simulations of cosmic ray transport through
turbulent magnetic fields to obtain a rigorous estimate for the hybrid viscosity in hot, 
two-temperature accretion disks. We find that the value of the Shakura-Sunyaev viscosity parameter
arising from this mechanism is in the range $0.01 \gtrsim \alpha_{\rm hyb} \gtrsim 0.02$ for 
weak to moderate Kraichnan turbulence ($0.1 \gtrsim \sigma^{2} \gtrsim 0.5$). 
We have thus demonstrated that hybrid viscosity is
a viable agent of steady state angular momentum transport in hot accretion disks.
In SBK96, we compared
the values of $\alpha_{\rm hyb}$ with those arising from pure magnetic viscosity $\alpha_{\rm mag}$.
In that paper, we had restricted ourselves to a considering a single value of 
$\xi \equiv \lambda_{\rm coh}/H$ at a time, 
and found that $\alpha_{\rm mag} = \xi^{2}$. This facilitated
a ready comparison between the two kinds of $\alpha$s. Here, however, we do not restrict ourselves to a single
value of $\xi$, and use the entire turbulent
spectrum to calculate $\alpha_{\rm hyb}$. We cannot, therefore, resort to such a direct comparison.
Simulations of the magnetorotational instability under conditions appropriate for
cold accretion disks suggest that $\alpha_{\rm mag} \sim 0.01$, but it is not clear what role pure magnetic
viscosity will play in hot accretion disks. The answer will depend crucially on the value of the plasma
$\beta$ attained in the steady state. 
\section{Acknowledgements}

\end{document}